\documentstyle[12pt,epsfig]{article}

\input epsf

\newcommand{\be}{\begin{equation}}
\newcommand{\ee}{\end{equation}}
\newcommand{\ba}{\begin{array}}
\newcommand{\ea}{\end{array}}
\newcommand{\baa}{\begin{eqnarray}}
\newcommand{\btab}{\begin{tabular}}
\newcommand{\etab}{\end{tabular}}
\newcommand{\eaa}{\end{eqnarray}}

\newcommand{\ci}[1]{\cite{#1}}

\newcommand{\lab}[1]{\label{#1}}

\newcommand{\edd}{\end{document}}

\newcommand{\alf}{\ifmmode\alpha \else$\alpha \ $\fi}
\newcommand{\bt}{\ifmmode\beta \else$\beta \ $\fi}
\newcommand{\gm}{\ifmmode\gamma \else$\gamma \ $\fi}
\newcommand{\Dl}{\ifmmode\Delta \else$\Delta \ $\fi}
\newcommand{\eps}{\ifmmode\varepsilon \else$\varepsilon \ $\fi}
\newcommand{\dl}{\ifmmode\delta \else$\delta \ $\fi}
\newcommand{\et}{\ifmmode\eta \else$\eta \ $\fi}
\newcommand{\vphi}{\ifmmode\varphi \else$\varphi \ $\fi}
\newcommand{\om}{\ifmmode\omega \else$\omega \ $\fi}
\newcommand{\pl}{\ifmmode\partial \else$\partial \ $\fi}
\newcommand{\ps}{\ifmmode\psi \else$\psi \ $\fi}
\newcommand{\sg}{\ifmmode\sigma \else$\sigma \ $\fi}
\newcommand{\phf}{\ifmmode\varphi^4 \else$\varphi^4 \ $\fi}
\newcommand{\Lam}{\ifmmode\Lambda \else$\Lambda$\fi}

\newcommand{\ppp}[1]{%
        \setbox0=\hbox{#1}%
        \kern-.02em\copy0\kern-\wd0
        \kern+.04em\copy0\kern-\wd0
        \kern-.02em\raise.0217em\box0}

\newcommand{\lsim}{
 \mathrel{\setbox0=\hbox{$<$}\raise0.6ex\copy0\kern-\wd0
 \lower0.65ex\hbox{$\sim$}}}

\newcommand{\gsim}{
 \mathrel{\setbox0=\hbox{$>$}\raise0.6ex\copy0\kern-\wd0
 \lower0.65ex\hbox{$\sim$}}}

\newcommand{\PRD}[3]{Phys.\ Rev.\ D {\bf {#1}}, {#2} ({#3})}

\newcommand{\PRL}[3]{Phys.\ Rev.\ Lett.\ {\bf {#1}}, {#2} ({#3})}

\newcommand{\PLB}[3]{Phys.\ Lett.\ B {\bf {#1}}, {#2} ({#3})}

\begin{document}

\setlength{\unitlength}{1mm}
\textwidth 16.0 true cm
\textheight 21.7 true cm
\headheight 0 cm
\headsep 0 cm
\oddsidemargin 0.10 true in
%
%

\begin{titlepage}
\renewcommand{\thefootnote}{\fnsymbol{footnote}}
\makebox[2cm]{}\\[-1in]
\begin{flushright}
\begin{tabular}{l}
TUM/T39-99-20\\
RUB-TPII-11/99\\
\end{tabular}
\end{flushright}
\vskip0.4cm
\begin{center}
  {\Large\bf
NLO corrections
and contribution of a tensor gluon operator to the process
$\gamma^*\gamma\rightarrow \pi\pi$ }


\vspace{0.7cm}

N. Kivel\footnote{Alexander von Humboldt fellow}$^{a,b}$,
L. Mankiewicz$^{a,c}$, M.V. Polyakov$^{b,d}$

\vspace{0.5cm}

\begin{center}

{\em$^a$Physik Department, Technische Universit\"{a}t M\"{u}nchen,
D-85747 Garching, Germany}

{\em $^b$Petersburg Nuclear Physics Institute,
  188350, Gatchina, Russia
}

{\em $^c$ N. Copernicus Astronomical Center, ul. Bartycka 18,
PL--00-716 Warsaw, Poland}

{\em $^d$ Institut f\"ur Theoretische Physik II, Ruhr-Universit\"at Bochum,
D-44780 Bochum, Germany}

\end{center}

\vspace{0.8cm}

{\em \today}

\vspace{0.8cm}

\centerline{\bf Abstract}
\begin{center}
\begin{minipage}{15cm}
We have calculated the NLO corrections for the leading twist amplitude of the
reaction $\gamma^*\gamma\to\pi\pi$ at large virtuality
of the photon and at $W^2\ll Q^2$.
With a simple model for the two-pion
quark and gluon distribution amplitudes we have estimated that the NLO
effects may reduce the leading-order amplitude by as much as 30\%.
We have found that besides the usual twist-2 quark and gluon operators the
NLO amplitude receives contribution from a
distribution associated with twist-2 tensor gluon operator. The new
gluon distribution can be investigated by studying the azimuthal
asymmetry of the cross section
for $e+\gamma\to e'+\pi\pi$.

\end{minipage}
\end{center}

\vspace{0.2cm}
{\em Submitted to Physics Letters B}
\end{center}
\end{titlepage}
\setcounter{footnote}{0}

\newpage

\noindent {\bf Introduction}\\

Investigation of the reaction $\gamma^*\gamma\to\pi\pi$ at large photon
virtuality but small invariant mass of the two-pion system ($W^2\ll
Q^2$) opens a new way to probe partonic structure of the pion
\cite{Pir98,Muller}. Under these conditions the corresponding amplitude
can be factorized
\cite{Pir98,Freund} into perturbatively calculable coefficient
functions
and nonperturbative soft matrix elements -- two-pion distribution
amplitudes ($2\pi$DA). These nonperturbative matrix elements contain
rich information about the partonic structure of the pion.

An interesting feature of the reaction $\gamma^*\gamma\to\pi\pi$
is that it is related by crossing to the deeply virtual Compton
scattering (DVCS) on a pion target. In particular, the $2\pi$DA's
which occur in the description of the  $\gamma^*\gamma\to\pi\pi$
leading-twist amplitude
can be related to the corresponding quark and gluon skewed distributions
in the pion \cite{MVP98,PW99}. As we shall discuss below, the
coefficient functions
in the reactions $\gamma^*\gamma\to\pi\pi$ and
$\gamma^*\pi \to \gamma\pi$ are also related
to each other by crossing.

{}From another point of view, the reaction $\gamma^*\gamma\to\pi\pi$ can
be considered as a minimal generalization of the transition form factor
$\gamma^*\gamma\to\pi^0$.  Due to the different isospin of the final
hadronic state these two processes are complementary to each
other -- the former probes isosinglet quark and gluon structure of the pion
while the latter is sensitive to the quarks in the isovector state.

The lowest order QCD contribution to the reaction
$\gamma^*\gamma\to\pi\pi$ has been recently analyzed in
Ref. \cite{Pir98}.  In this Letter we extend this analysis to the next
to leading order (NLO) in $\alpha_s(Q^2)$. The leading-order,
leading-twist amplitude accommodates only collisions of two photons
with equal helicities, related by crossing to the photon
helicity-conserving DVCS on a pion.  At the NLO there is a new
contribution from collisions of photons with opposite helicities,
related to photon helicity-flip contribution to DVCS. In terms of the
Operator Product Expansion this amplitude singles out a twist-2 tensor
gluon operator which cannot be studied in deep-inelastic scattering
(DIS) on a nucleon target. It was introduced by Jaffe and Manohar
\cite{JaffMan} in their studies of DIS on a spin-1 target. Its
relation to DVCS on a nucleon and to the current fragmentation in
DIS have been considered in Refs.~\cite{JiHoodbhoy} and \cite{SST},
respectively. As we discuss below it can be also studied via
azimuthal asymmetries in the hard photoproduction of a pion pair.

\vspace{1cm}

\noindent{\bf General definitions}\\

Reaction $\gamma^* \gamma \to \pi \pi$ is schematically depicted in Fig.1.
The kinematics can be conveniently described in terms of a pair of light-like
vectors $n,n^*$ which obey
\be
\ba{l}
n^2 = n^{*2} = 0, \, n \cdot n^* = 2,\\
n = (1,0,0,-1), \, n^* = (1,0,0,1)\,
\ea
\label{nvectors}
\ee
and define longitudinal directions. Vectors $n,n^*$ can be also used to
define the
metrics tensor in the transverse space
\begin{equation}
(-g^{\mu\nu})_\perp =
\left( \frac{n^\mu n^{*\nu} + n^\nu n^{*\mu}}{n \cdot n^*}
  - g^{\mu\nu}\right) \,
\label{transverse_metrics}
\end{equation}
such that $(-g^{\mu\nu})_\perp n_\mu = (-g^{\mu\nu})_\perp n^*_\mu = 0$.
$$
\epsfig{file=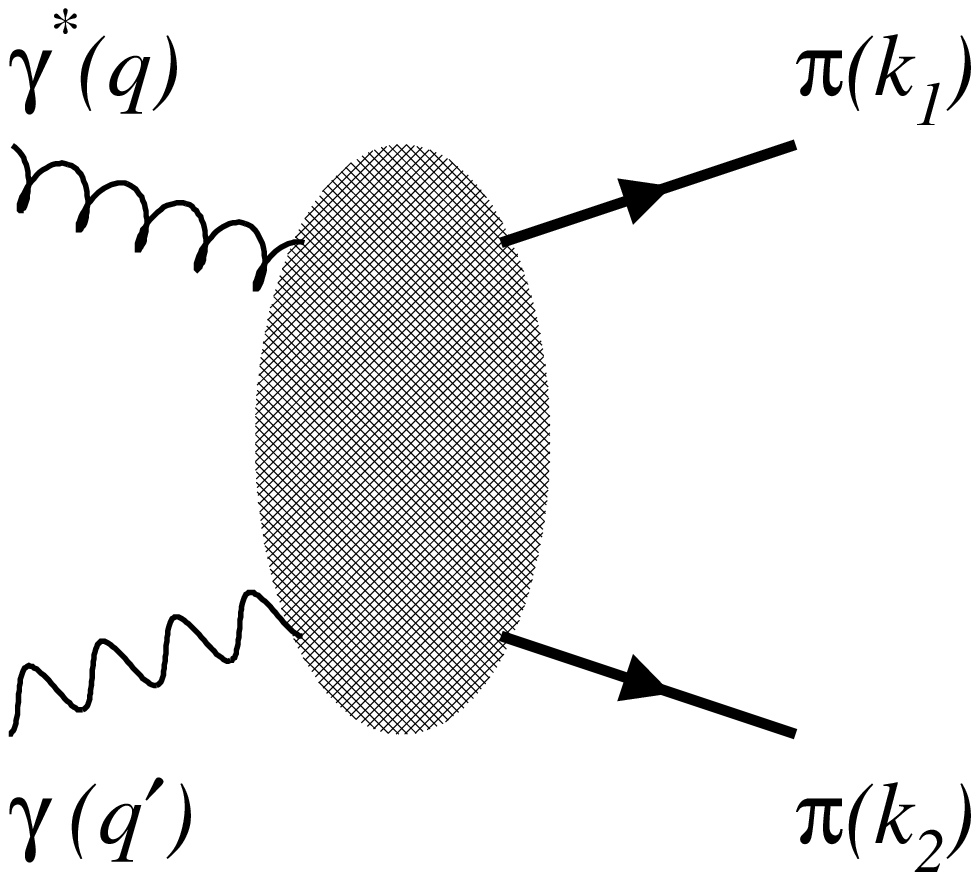,height=4cm,width=4cm}
$$
\centerline{Fig. 1}
\vskip0.5cm
Initial and final states momenta can be conveniently decomposed as
\be
\ba{l}
\displaystyle
 q = \frac{-Q}{2}\, n + \frac{Q}{2}\, n^*, \, q^2 = - Q^2\, \hspace*{8mm}
q^\prime = (\frac{Q}{2} + \frac{W^2}{2 Q})\, n ,\mskip10mu q^{\prime 2} = 0,
\\[4mm]\displaystyle
 P = q + q^\prime = \frac Q2 \, n^* + \frac{W^2}{2 Q}\,  n, \, \, P^2 = W^2
\\[4mm] \displaystyle
k_1 =\frac12 \zeta Q \, n^* +
\frac12 \frac{{\bar \zeta}W^2}{ Q}\, n + K_\perp,
\mskip10mu
k_2 = \frac12{\bar \zeta}Q\, n^* +\frac12 \frac{\zeta W^2}{Q}\, n
- K_\perp
\ea
\label{lcexpansion}
\ee
Here we have introduced the total momentum $P$ of the pion pair, $P = k_1 +
k_2$, and the momentum fractions $\zeta=(k_1 \cdot n)/(P \cdot n),
{\bar \zeta}=(k_2 \cdot n)/(P \cdot n), \zeta + {\bar \zeta} = 1$, which
describe
the distribution of
longitudinal momentum between two pions. Alternatively,
$$2 \zeta - 1 = \beta \cos\theta_{\rm cm}\, ,$$
where $\theta_{\rm cm}$ is the polar
angle of the pion momentum in the CM frame with respect to
the direction of the total momentum $\vec P$ and $\beta$ is the velocity of
produced pions in centre of mass frame
\begin{eqnarray}
\nonumber
\beta=\sqrt{1-\frac{4 m_\pi^2}{W^2}}\, .
\end{eqnarray}

The amplitude of
hard photoproduction of two pions
is defined by the following matrix element
between vacuum and two pions state \ci{Pir98}:
\be
T^{\mu\nu} = i \int d^4x e^{-ix\cdot {\bar q}}
\langle 2 \pi (P)| T J^\mu(x/2) J^\nu(-x/2) | 0 \rangle\, ,
\mskip10mu {\bar q}=\frac12(q+q')\, ,
\label{basic_amplitude}
\ee
where $J^\mu(x)$ denotes the quark electromagnetic current.  Hard
photoproduction corresponds to the limit
$Q^2 \gg \Lambda_{\rm QCD}^2, \, W^2$
where the amplitude (\ref{basic_amplitude}) can be represented as an expansion
in terms of powers of $1/Q$. According to the factorization theorem
\cite{Pir98,Freund}
the leading twist term in the expansion can be written as a convolution of
hard and soft blocks, see Fig.2. The coefficient functions, defined by hard
blocks, can be calculated from appropriate partonic subprocesses
$\gamma^*+\gamma\rightarrow \bar{q}+q$ or $\gamma^*+\gamma\rightarrow g+g$.
$$
\ba{l}
\epsfig{file=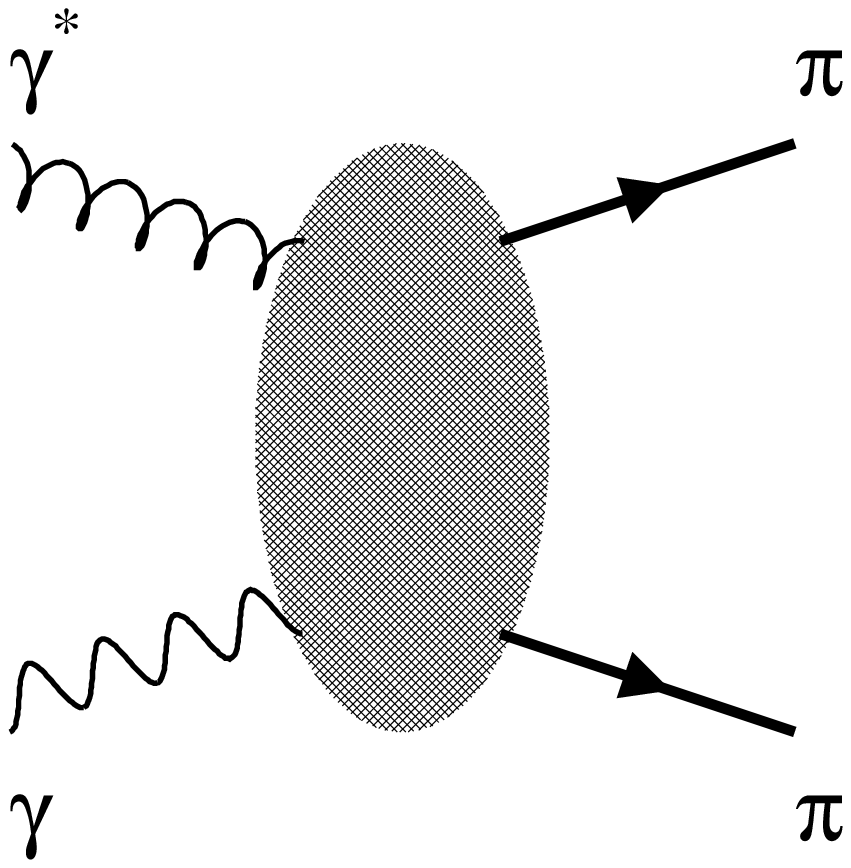,height=3.5 cm,width=3.5cm}
\ea
=
\ba{l}
\epsfig{file=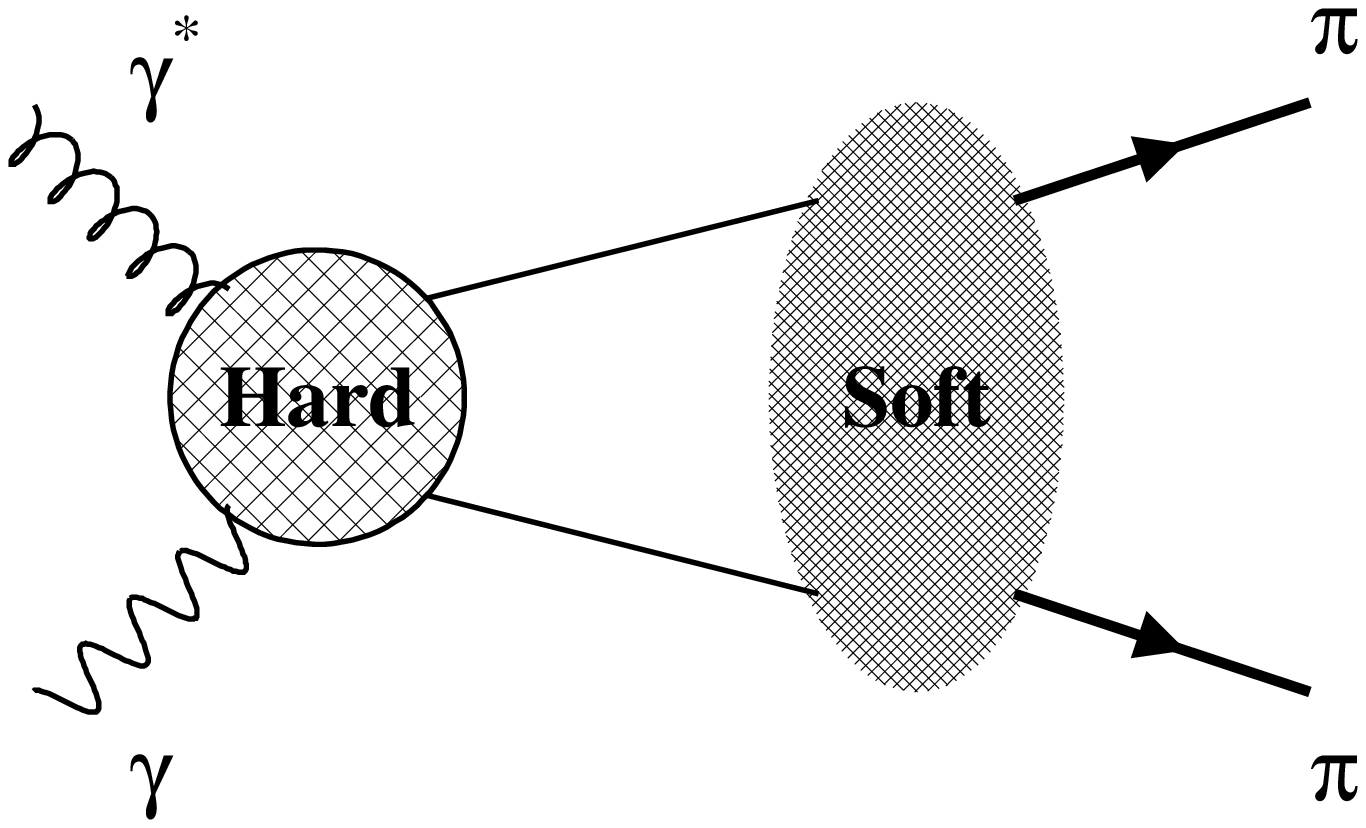,height=3.5cm,width=5.4cm}
\ea
+ \mbox{~power corrections}
$$
{}\\[-0.1cm]
\centerline{Fig.2 Factorization of hard and soft contributions to the
  amplitude (\ref{basic_amplitude}).}
\\[1mm]

Soft blocks represent two-pion distribution amplitudes (2$\pi$DA), or
probability amplitudes for a conversion of a pair of
collinear partons (quarks or gluons) into the final pion pair. The quark
2$\pi$DA is defined as a matrix element of a light-cone
quark string operator:
\be
\langle \pi\pi(P) | {\bar q}(x) {\hat n} q(0) |0\rangle_{x^2 = 0}
= P\cdot n \int_0^1 dz \, \Phi^q(z,\zeta,W^2) \, e^{izP\cdot x} \, ,
\lab{quarkDA}
\ee
with ${\hat n} = \gamma^\mu n_\mu$. Here and below the appropriate 
path-ordered exponential between points $x$ and $0$ is tacitely assumed.
Due to the charge-conjugation invariance
the
amplitude of the process $\gamma^*\gamma\to\pi\pi$ depends
only on the quark $2\pi$DA with pions in isospin zero state:
\be
\Phi^Q=\frac{1}{N_f}\sum_q \Phi^q \, .
\ee

Beyond the leading order, pions can originate from a pair of collinear gluons
as well. Here two different gluon 2$\pi$DA's enter. The
distribution amplitude $\Phi^G (z, \zeta, W^2)$ defined as
\be
\langle \pi\pi(P) | n^\mu n^\nu \,
G^{A\,\mu}_{\ \ \ \alpha}(x) G^{A\,\alpha\nu}(0)
|0\rangle_{x^2 = 0}
=(n\cdot P)^2\int_0^1 dz\ 
\Phi^G (z, \zeta, W^2 )\ e^{i z (P\cdot  x)}\; ,
\label{definitionG}
\ee
describes the situation when pions arise from two gluons with opposite
helicities, see below. As we shall discuss in the following, it can be related
by crossing to the ordinary gluon distribution in a pion. The second
distribution amplitude $\Phi^{TG} (z, \zeta, W^2)$ can be defined as
\be
\ba{l}
\displaystyle
\langle \pi\pi(P) | n^\mu n^\nu \,
G^{A}_{\mu\alpha_\perp}(x) G^{A}_{\nu\beta_\perp}(0)
|0\rangle_{x^2 = 0}
= \\[6mm]
\displaystyle
\mskip30mu
=(n\cdot P)^2
\frac{K_{\alpha_\perp} K_{\beta_\perp}-\frac 12
(-g_{\alpha_\perp \beta_\perp}) K_\perp^2}{W^2}
\int_0^1 dz\ 
\Phi^{TG} (z, \zeta, W^2)\ e^{i z (P\cdot  x)} \, .
\ea
\label{definitionTG}
\ee
Here the indices $\alpha_\perp \beta_\perp$ are transverse with respect to
vectors $n,n^*$ and occur in a symmetric, traceless combination. $\Phi^{TG}$
represents fragmentation of two gluons with equal helicities into the pion
pair. It is related by crossing to the gluon helicity-flip skewed pion
distribution introduced by Ji and Hoodbhoy \cite{JiHoodbhoy}.

\vspace{1cm}

\noindent{\bf Properties of two-pion distribution amplitudes}\\

Let us now briefly discuss properties of
two-pion distribution amplitudes.
Quark $2\pi$DA's were discussed extensively in Ref.~\cite{MVP98}.
The gluon distribution amplitude $\Phi^G (z, \zeta, W^2)$
(\ref{definitionG}) is constrained by soft pion theorem and crossing
relations.  The soft pion theorem for $\Phi^G (z, \zeta, W^2)$ can be
obtained in the same way as for the quark distribution \cite{MVP98}. It reads:
\be
\Phi^G (z, \zeta= 0, W^2= 0 )=
\Phi^G (z, \zeta= 1, W^2= 0 )=0\, .
\label{sptg}
\ee

Following \cite{MVP98}, we decompose both quark and gluon $2\pi$DA's
in conformal and partial waves.
For quark isosinglet
$2\pi$DA the decomposition reads
\be
\Phi^Q(z, \zeta, W^2 )=6z(1-z)
\sum_{\scriptstyle n=1 \atop \scriptstyle {\rm odd}}^{\infty}
\sum_{\scriptstyle l=0 \atop \scriptstyle {\rm even}}^{n+1} B_{nl}(W) C_n^{3/2}(2 z-1)
P_l(2\zeta-1),
\label{razhlq}
\ee
while for gluon $2\pi$DA's we have
\be
\Phi^G(z, \zeta, W^2 )=
30\ z^2(1-z)^2
\sum_{\scriptstyle n=0 \atop \scriptstyle {\rm even}}^\infty
\sum_{\scriptstyle l=0 \atop \scriptstyle {\rm even}}^{n+2}
A^G_{nl}(W) C^{5/2}_n(2 z-1) P_l(2\zeta-1)\, ,
\label{razlg}
\ee
\be
\Phi^{TG}(z, \zeta, W^2 )=
10\ z^2(1-z)^2
\sum_{\scriptstyle n=0 \atop \scriptstyle {\rm even}}^\infty
\sum_{\scriptstyle l=2 \atop \scriptstyle {\rm even}}^{n+2}
A^{TG}_{nl}(W) C^{5/2}_n(2 z-1) P_{l}^{\, \prime \prime}(2\zeta-1)\, .
\label{razltg}
\ee 
Here $P_l(x)$ is a Legendre polynomial and $P_{l}^{\, \prime
  \prime}(x)=d^2 P_l(x)/dx^2$. With this choice of the polynomial basis for the
  $\zeta$-expansion the above decomposition is
directly related to the partial-wave expansion of the resulting two-pion
production amplitude.

Additional constraints for quark and gluon $2\pi$DA $\Phi^G$
are provided by the crossing relations between $2\pi$DA's
and known parton distributions in the pion.
The derivation of such relations for quark $2\pi$DA's
see \cite{MVP98,PW99}. For gluon $2\pi$DA the derivation
is analogous. The crossing relations have the form:
\be
\ba{l}
\displaystyle
B_{N-1,N}(0)=
\frac 23\ \frac{2N+1}{N+1} \int_{0}^1 dx\ x^{N-1}
\frac {1}{N_f}\sum_q(q_\pi(x)+\bar q_\pi(x))\; ,
\\[8mm] \displaystyle
A_{N-2,N}^G(0)=\frac{4}{5}\
\frac{2N+1}{(N+1)(N+2)}
\int_{0}^1 dx\ x^{N-1}
g_\pi(x)\; ,
\ea
\label{crossing}
\ee
where $q_\pi(x)$, $\bar q_\pi(x)$ and $g_\pi(x)$ are usual quark, antiquark
and gluon distributions in the pion.
Using these relations one
can easily derive the normalization of $2\pi$DA's
at $W^2=0$:
\be
\ba{l}
\displaystyle
\int_0^1 dz (2 z-1)\Phi^Q(z, \zeta, W^2=0 )
=-\frac{4}{N_f}\ M_2^{Q}\zeta(1-\zeta) \, ,
\\[4mm] \displaystyle
\int_0^1 dz\ \Phi^G (z, \zeta, W^2=0)=
-2\ M_2^G\zeta(1-\zeta) \, .
\ea
\ee
Here $M_2^Q$ and $M_2^G$ are momentum fractions carried
respectively by quarks and gluons in the pion.

Soft pion theorems and crossing relations constrain
$\Phi^Q(z,\zeta,W^2)$ and $\Phi^G (z, \zeta, W^2 )$
only at $W^2=0$. For higher $W^2$ one can apply dispersion
relation analysis of Ref.~\cite{MVP98}, see this paper
for more details.

Under evolution $\Phi^G$
and $\Phi^Q$ mix with each other \cite{mix}.
For asymptotically large $Q^2$ one obtains (taking into account
soft pion theorem (\ref{sptg}))
\be
\ba{l}
\displaystyle
\Phi^G_{asy} (z, \zeta, W^2=0 )=
- \frac{240\ C_F}{N_f+4 C_F}\ z^2(1-z)^2 \ \zeta(1-\zeta)\, ,
\\[4mm]  \displaystyle
\Phi^Q_{asy} (z, \zeta, W^2 =0 )=
-\frac{120}{N_f+4 C_F}\ z(1-z)\ (2 z-1) \ \zeta(1-\zeta)\, .
\ea
\label{asymptotic}
\ee
The distribution amplitude $\Phi^{TG}$
does not mix with quarks under evolution and
vanishes at asymptotically large $Q^2$.
Taking into account contribution corresponding to the smallest anomalous
dimension leads to the following ``preasymptotic'' expression
\be
\Phi^{TG}_{asy} (z, \zeta, W^2=0 )=
\Biggl(\frac{\alpha_s(Q^2)}{\alpha_s(\mu^2)}\Biggr)^{\gamma^{TG}/\beta_0}
30\ z^2(1-z)^2\ A_{02}^{TG}(0)
\, ,
\label{preasymptotic}
\ee
where $A_{02}^{TG}(0)$ is
unknown constant depending on the normalization point $\mu$.
This constant is related to matrix element of local operator
$G^{A}_{\mu\alpha_\perp}(0) G^{A}_{\nu\beta_\perp}(0)$ in a pion
and generically can be of order unity.
In eq.~(\ref{preasymptotic}) dependence on the scale $Q^2$ is shown
explicitly. The corresponding
anomalous dimension $\gamma^{TG}$ was calculated in Ref.~\cite{JiHoodbhoy}:
\be
\gamma^{TG}=\frac{7\ N_c}{3}+ \frac{2\ N_f}{3}\, .
\ee

\vspace{1cm}

\noindent{\bf Helicity structure of hard partonic amplitudes}\\

Let us consider the hard amplitudes
$\gamma^*+\gamma\rightarrow\bar{q}q$ and $\gamma^*+\gamma\rightarrow g+g$.
Let $\lambda^*, \lambda$ denote the respective helicities of the
virtual and real photon and $\lambda_1, \lambda_2$ the helicities of produced
partons, quarks or gluons. Due to parity invariance, the helicity amplitude
 $M_{\lambda^*\lambda,\lambda_1\lambda_2}$ must obey
  $M_{\lambda^*\lambda,\lambda_1\lambda_2}=
M_{-\lambda^*-\lambda,-\lambda_1-\lambda_2}$. Because hard scattering
occurs
collinearly, conservation of the angular momentum along the
collision axis results in the helicity conservation
$\lambda^*-\lambda=\lambda_1+\lambda_2$. For the case when photons create a
quark-antiquark pair
one finds the following amplitudes
\be
\ba{l}
L_z =0:\mskip20mu
M_{11,\frac12,-\frac12} \mbox{ or }\mskip10mu  M_{11,-\frac12,\frac12}
\\[3mm]
L_z=\pm1:\mskip20mu
M_{0-1,\frac12,\frac12} \mbox{ or }\mskip10mu  M_{01,-\frac12,-\frac12}
\ea
\ee
Note, however that the $L_z=\pm1$ amplitudes require a helicity flip along the
fermion line and therefore vanish to the leading twist accuracy. It is easy to
show that the quark
$2\pi$DA $\Phi^Q$ (\ref{quarkDA}) describes a subsequent evolution of a
collinear quark-antiquark pair with $L_z =0$ into two pions. For the case when
two gluons emerge from the hard collision one finds
\be
\ba{l}
L_z=0:\mskip20mu
M_{11,1,-1} \mbox{ or }\mskip10mu  M_{11,-1,1}
\\[3mm]
L_z=\pm2:\mskip20mu
M_{1-1,11}
\ea
\lab{helampitude}
\ee
A short calculation shows that the evolution of two collinear gluons with
$L_z=0$ and $L_z=\pm 2$ into two pions is described by the distribution
amplitudes $\Phi^G$ and $\Phi^{TG}$, respectively. Note that the
latter results in a
nontrivial dependence of the cross section on
azimuthal angle $\varphi$ between
the lepton scattering plane and the pion plane. As it follows, one can extract
information about tensor gluon distribution $\Phi^{TG}(z,\zeta,W^2)$
from the corresponding differential cross section.

\vspace{0.8cm}

\noindent{\bf The leading twist $\gamma^* \gamma \to \pi \pi$ amplitude to the
  NLO accuracy}\\

Here we present the NLO results for the leading-twist
 contribution to the amplitude (\ref{basic_amplitude}). First, we recall the
 LO contribution discussed in \cite{Pir98}.
The Born graphs contributing to the LO coefficient function
are shown in Fig. 3.
\\[-10mm]
$$
\ba{l}
\epsfig{file=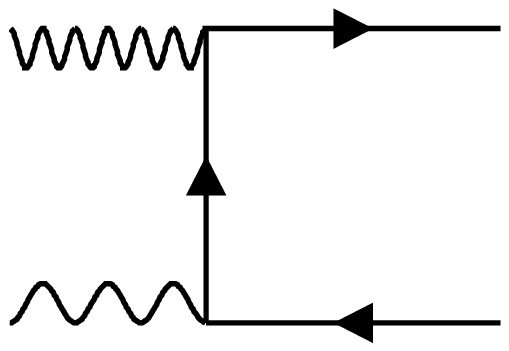,height=3cm,width=3cm}
\ea
\mskip-40mu
+\mskip10mu
\mbox{crossed}
$$
{}\\[-10mm]
\centerline{Fig. 3 LO contributions to the coefficient function}
\\[0.5cm]

To this accuracy the colliding photons can produce a
quark-antiquark pair only.  A direct calculation yields \cite{Pir98}
\be
T^{\mu\nu} =  \frac i2 \left( \sum e_q^2\right)\, (-g^{\mu\nu})_T \int_0^1 dz\,
\Phi^Q(z,\zeta,W^2) \left( \frac{1}{{\bar z}} - \frac{1}{z}\right)
\ee
Note that up to QCD evolution effects the amplitude scales as a function of
$Q^2$ at fixed invariant mass $W^2$ and $\zeta$. One-loop corrections to
quark and gluon coefficient functions arise from diagrams depicted in Figs.4
and 5.
\\[-10mm]
$$
\ba{l}
\mskip-10mu
\ba{l}
\epsfig{file=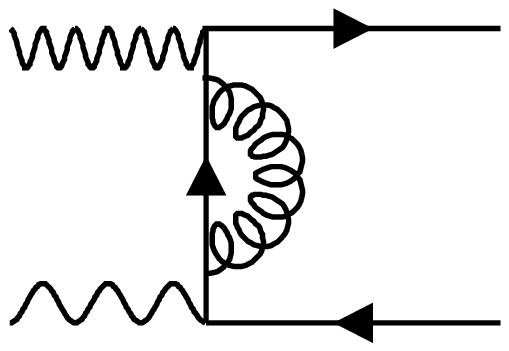,height=3.5cm,width=3.5cm}
\ea
\mskip-55mu
+
\mskip-10mu
\ba{l}
\epsfig{file=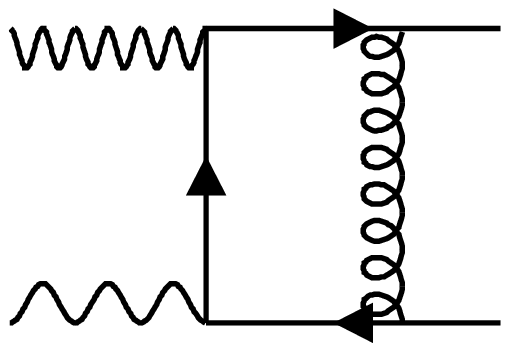,height=3.5 cm,width=3.5cm}
\ea
\mskip-45mu
+
\mskip-10mu
\ba{l}
\epsfig{file=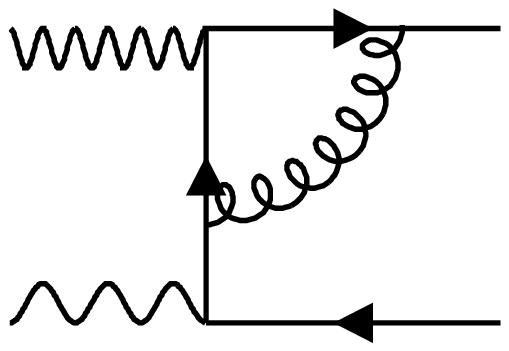,height=3.5 cm,width=3.5cm}
\ea
\mskip-55mu
+
\mskip-10mu
\ba{l}
\epsfig{file=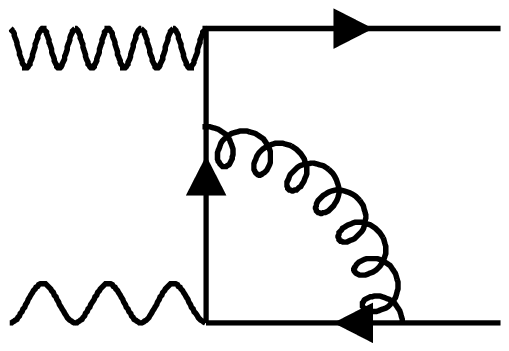,height=3.5 cm,width=3.5cm}
\ea
\mskip-55mu
+\mskip15mu \mbox{crossed}
\ea
$$
{}\\[-10mm]
\centerline{Fig4. One-loop contributions to the quark coefficient function }
\\[6mm]
The final result for amplitudes corresponding to the hard subprocess with
$L_z = 0$ can be written as:
\be
\ba{l}
\displaystyle
T^{\mu\nu}_{L_z=0}
=
\frac i2 \left( \sum e_q^2 \right) (-g^{\mu\nu})_T \int_0^1 dz\,
\Phi^Q(z,\zeta,W^2) \left[C^0_q(z)+
{\textstyle \frac{\alf_S(Q^2)}{4\pi}}
C^1_q(z)\right]
\\[8mm]
\displaystyle
\hspace*{10mm}
- \frac{i}{2} \left(\sum e_q^2\right) (-g^{\mu\nu})_T
\int_0^1 dz\,
\Phi^G(z,\zeta,W^2)\left[
{\textstyle \frac{\alf_S(Q^2)}{4\pi}} C^1_g(z)
\right] \, .
\ea
\label{NLOLz0}
\ee
%

As discussed in the previous section, gluons give rise to the hard amplitude
with $L_z = \pm 2$ which results in the contribution
\be
\ba{l}
\displaystyle
T^{\mu\nu}_{L_z = \pm 2}=
- i \left( \sum e_q^2 \right)
 \frac{K_\perp^{(\mu} K_\perp^{\nu)}}{W^2} \,
\int_0^1 dz \, \Phi^{TG}(z,\zeta,W^2)
\left[{\textstyle\frac{\alf_S(Q^2)}{4\pi}}C^1_{T}(z)\right]\, .
\ea
\lab{NLOLz2}
\ee
Here $K_\perp^{(\mu} K_\perp^{\nu)}=
K_\perp^\mu K_\perp^\nu- 1/2 (-g^{\mu\nu})_\perp K_\perp^2 $ and $K_\perp$ is
the transverse component of the pions momenta (\ref{lcexpansion}).
\\[-10mm]
$$
\mskip-10mu
\ba{l}
\epsfig{file=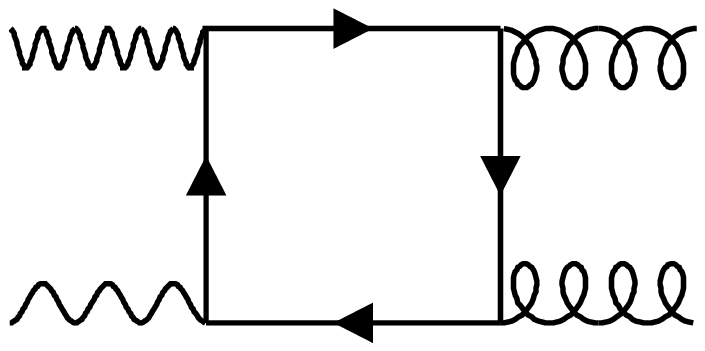,height=3.5 cm,width=3.5cm}
\ea
\mskip-20mu
+
\mskip-10mu
\ba{l}
\epsfig{file=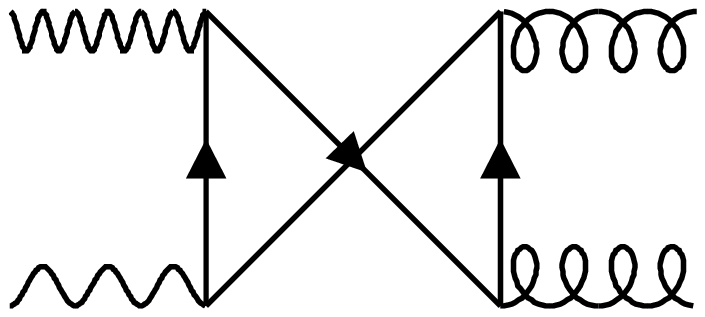,height=3.5cm,width=3.5cm}
\ea
\mskip-20mu
+
\mskip-10mu
\ba{l}
\epsfig{file=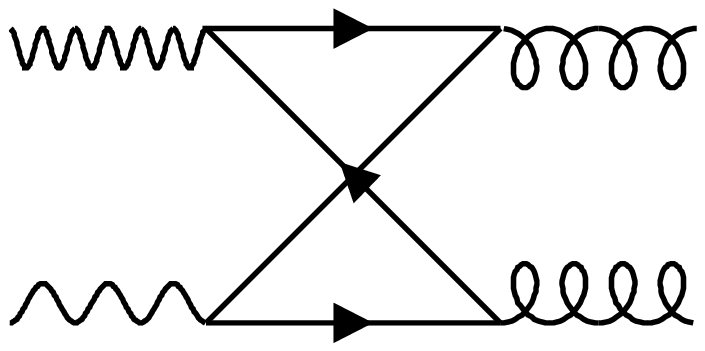,height=3.5 cm,width=3.5cm}
\ea
\mskip-20mu
+\mskip20mu \mbox{crossed}
$$
{}\\[-10mm]
\centerline{Fig.5 Diagrams for gluon coefficient functions}
\\[2mm]
Below we present explicit formulae for the coefficient functions $C^1_q,C^1_g$
and $C^1_T$ introduced in (\ref{NLOLz0},\ref{NLOLz2}). Note that they can be
obtained either by a direct calculation, or in a much simpler way
by using the
crossing symmetry of hard partonic amplitudes which relates them to the
previously computed NLO coefficient functions for the DVCS process
\cite{Belxx,JiOsb98,Manketal98}.
\be
\ba{l}
\displaystyle
C^1_q(z)=C_F\left\{ \frac{\ln^2(1-z)}{z(1-z)}-\frac{\ln^2{z}}{z(1-z)}+
\frac{\ln^2{z}}{(1-z)}-\frac{\ln^2{(1-z)}}{z}+\right.
\\[5mm]\displaystyle \mskip50mu
\left.+3\frac{\ln{z}}{(1-z)}-
3\frac{\ln{(1-z)}}{z}+\frac9z-\frac9{(1-z)}\right\}\, ,
\\[8mm]
\displaystyle
C^1_g(z)=\frac1{z^2(1-z)^2}\left[ z^2\ln^2z+(1-z)^2\ln^2(1-z)+\right.
\\[5mm]
\displaystyle\mskip50mu
\left.+ 2z(1-z)\ln[z(1-z)]-
4z\ln z-4(1-z)\ln(1-z)\right]\, ,
\\[4mm]
\displaystyle
C^1_T(z)= \frac{2}{z(1-z)}\, .
\ea
\lab{coeffun}
\ee

To get the complete NLO expression one has to include corresponding
corrections to the $\Phi^Q$ and $\Phi^G$ distribution
amplitudes. Previously, the NLO corrections to the distribution
amplitudes in the $\overline{MS}$ scheme have been found numerically small
\ci{Bel98,Rad86,DMull95}. In the following numerical analysis we have
assumed that their effects can be neglected as compared with
uncertainties introduced by assumptions necessary to model the
distribution amplitudes themselves.

\vspace{1cm}

\noindent{\bf Numerical results}\\

Here we present numerical estimates of the NLO effects in the leading-twist
amplitude for $\gamma^* \gamma \to \pi \pi$ process. For this purpose, we
assume that the corresponding $2\pi$DA's are not far from their
asymptotic, respectively preasymptotic forms, see
eqs.~(\ref{asymptotic},\ref{preasymptotic}).
To continue $2\pi$DA's from the point
$W^2=0$ to the higher $W^2$ values one applies the Watson
final state interaction theorem \cite{Watson} and Omn\`es solution
\cite{omnes}
to corresponding dispersion relations, see Ref.~\cite{MVP98} for details.
With these assumptions, the $L_z = 0$ amplitude
reads:
\baa
T^{\mu\nu}_{L_z=0}&=&\frac i 2 (\sum_q e_q^2 )\, (-g^{\mu\nu})_\perp
\ \frac{-40}{N_f+4 C_F}
\Biggl\{
 1 -\frac {87}{9}\ \frac{\alpha_s(Q^2)}{4\pi}
\  C_F
\Biggr\}\nonumber \\
&\times& \Biggl[
\frac{3C-\beta^2}{12}\ f_0(W)\ P_0(\cos\theta_{\rm cm}) -
\frac{\beta^2}{6} \ f_2(W) \ P_2(\cos\theta_{\rm cm})
\Biggr]
\, .
\label{nonflip}
\eaa
Let us note that the asymptotic $2\pi$DA's
eqs.~(\ref{asymptotic},\ref{preasymptotic}) allow production of
the pions only in S- and D-waves. Deviations of the $2\pi$DA's
from their asymptotic form result in appearence of higher waves
of produced pions.

The functions $f_{0,2}(W)$, so-called Omn\`es functions,  can be
related to $\pi\pi$ phase shifts $\delta_0^0(W)$ and
$\delta_2^0(W)$ using Watson theorem and dispersion
relations derived in \cite{MVP98}:

\be
f_l(W)=
\exp\biggl[i\delta_l^0(W)+
\frac{W^{2}}{\pi}
{\rm Re} \int_{4m_\pi^2}^\infty
ds \frac{\delta_l^0(s)}{s(s-W^2-i0)}
\biggr]\, .
\label{omnes}
\ee
The Omn\`es functions were analyzed in details in Ref.~\cite{Gasser}
and for the estimates of the cross sections one can use the results
of Ref.~\cite{Gasser}. The corresponding estimates will be given
elsewhere.

The constant $C$ in eq.~(\ref{nonflip}) plays a role of an integration
constant in 
the Omn\`es solution to the corresponding dispersion relations.
From the soft pion theorem it follows that $C=1+O(m_\pi^2)$.

Let us note that the expression (\ref{omnes})
is valid strictly speaking only in the elastic region
($4 m_\pi^2 \leq W^2 \leq 16 m_\pi^2$).
It should not be complicated to extend its range of applicability including
the contributions of higher intermediate states  (probably the most
important is the contribution of $K\bar K$) in the dispersion relations.
Note that the accuracy of the Omn\`es solution can be increased
if one has an independent information about, say, slopes of
functions $f_l(W)$ at small $W$. Such information can be provided by
effective low energy models of QCD \cite{PolWeiss99}.
For example, using the instanton model calculations
of $B_{nl}(W)$ at low energies \cite{MVP98} one finds the constant
$C$ in eq.~(\ref{nonflip}) to be equal to:
\be
\nonumber
C=1+b\ m_\pi^2 + O(m_\pi^4)\ {\rm \ with \ \ \ \ }
b\approx -1.7\ {\rm GeV}^{-2}\, .
\ee

The Omn\`es functions contain information about the isoscalar $\pi\pi$
resonances as well as about the nonresonant background. Due to the resonance
contributions $f_0(W)$ is enhanced
around $W=1$~GeV ($f_0(980)$ resonance)
whereas $f_2(W)$ has a peak at $W=1.275$~GeV ($f_2(1270)$ resonance).
Let us note that close to the threshold $W=2 m_\pi$ production
of pions occurs mostly in the $S-$wave. The corresponding Omn\`es
function $f_0(W)$ is fixed near the threshold by the
Chiral Perturbation Theory \cite{Gasser} to be:
\be
f_0(W)=1+
\frac{2 W^{2}-m_\pi^2}{32 \pi^2 f_\pi^2}
\biggl\{ \beta\ \log\biggl(\frac{1-\beta}{1+\beta}\biggr) +2+
i\pi\beta
\biggr\}+ \frac{W^2}{192\pi^2f_\pi^2}\, .
\label{omneschpt}
\ee

Numerically, from equation (\ref{nonflip}) it follows
that the NLO correction
reduce the LO $L_z = 0$ amplitude by about 30\% and that
the major NLO correction is due to the gluon $2\pi$DA.

With $\Phi^{TG}$ given by equation (\ref{preasymptotic}) the $L_z = \pm 2$
amplitude reads
\baa
\nonumber
T^{\mu\nu}_{L_z = \pm 2}&=&
-\frac i 2 \sum_q e_q^2 \ \frac{\alpha_s(Q^2)}{4\pi}
\frac{K^\mu_\perp K^\nu_\perp-\frac 12
(-g^{\mu\nu})_\perp K_\perp^2}{W^2}\\
&\times&20\
\Biggl(\frac{\alpha_s(Q^2)}{\alpha_s(\mu^2)}\Biggr)^{\gamma^{TG}/\beta_0}
\ A_{02}^{TG}(0)\ f_2(W)\, .
\eaa
Note that the lowest possible partial wave of produced pions
in this case is that with $l=2$. It implies that
the $L_z=\pm 2$ amplitude is strongly
enhanced around $W=1.275$~GeV due to contribution of $f_2(1270)$
resonance.

Studies of the NLO effects using models of $2\pi$DA's beyond the
asymptotic approximations (\ref{asymptotic},\ref{preasymptotic}) will be
published elsewhere (see
\cite{DIS99} for model predictions for the LO amplitude).

The interference of $L_z=0$ and $L_z=\pm 2$ amplitudes produces the
azimuthal asymmetries $\sim \cos 2\varphi$ 
in
the cross section of $e+\gamma\to e'+\pi\pi$ reactions, where
$\varphi$ is the angle between the lepton scattering plane and the pion plane,
respectively. 
As the
energy dependence of the $\pi\pi$ phase shifts is known from
low-energy experiments, the $W-$dependence of various asymmetries can
be to great extend fixed by the low-energy data. This makes the reaction
$\gamma^*\gamma\to\pi\pi$ one of the cleanest possible sources of
information about the tensor gluon distribution amplitude $\Phi^{TG}$.

The process of the interest $\gamma^*\gamma\to\pi\pi$ in the reaction
$e+\gamma\to e'+\pi\pi$ competes with electromagnetic background
process when the pion pair is created by a bremsstrahlung photon.
As the bremsstrahlung results in the pion pair being
produced in a $C-$odd state this contribution is absent for the $\pi^0\pi^0$
final state.
In addition, the interference of the background process with
$\gamma^*\gamma\to\pi^+\pi^-$ can be selected by various charge
asymmetries \cite{Pir98,Pir99b}.

To our best knowledge at present there are no dedicated measurements of the
process $\gamma^*\gamma\to\pi\pi$. To the contrary, the processes with
real photons were studied extensively, see e.g \cite{Edwards} and
\cite{Behrend} for
$\pi^0\pi^0$  and $\pi^+\pi^-$ channels, respectively.
Let us note that in the CLEO measurements of $\gamma \pi$ transition
formfactor at large $Q^2$ \cite{Savinov} the $\gamma^* \gamma\to \pi^0\pi^0$
were considered background events. We strongly recommend to reanalyze this
data in order to extract the $\gamma^* \gamma\to \pi^0\pi^0$
process.

To summarize, our estimates suggest that the NLO corrections to the
leading-twist $\gamma^*\gamma\to\pi\pi$ amplitude are not small and
therefore of large importance for analysis of an experimental
data. Moreover, we have shown that at the NLO accuracy the amplitude
contains a novel tensor gluon distribution amplitude, which can be studied
via azimuthal asymmetries in the cross-section.

\vspace{1cm}

\noindent
{\bf Acknowledgments:}

\noindent
This paper owes much to many comments and suggestions by Marcus
Diehl. We gratefully acknowledge his contribution.
N.K. would like to thank the Humboldt Foundation for financial
support. We acknowledge useful discussions with
K. Goeke, G. Piller, W. Weise and C. Weiss.
\\

\end{document}